\documentclass[12pt]{iopart}
\usepackage{iopams}
\usepackage{graphicx}
\usepackage{amssymb}

\usepackage{color}

\begin{document}

\title{Optimal operating conditions of an entangling two-transmon gate}
\author{Antonio D'Arrigo}
\ead{adarrigo@dmfci.unict.it}
\address{Dipartimento di Fisica e Astronomia,
Universit\`a di Catania,  c/o  Viale A. Doria 6, Ed. 10, 95125 Catania, Italy
and CNR - IMM - MATIS c/o DFA Via Santa Sofia 64, 95123 Catania, Italy}
\author{Elisabetta Paladino}
\ead{epaladino@dmfci.unict.it}
\address{Dipartimento di Fisica e Astronomia,
Universit\`a di Catania,  c/o  Viale A. Doria 6, Ed. 10, 95125 Catania, Italy
and CNR - IMM - MATIS c/o DFA Via Santa Sofia 64, 95123, Catania, Italy}

\begin{abstract}  
We identify optimal operating conditions of an entangling two-qubit
gate realized by a capacitive coupling of two superconducting charge qubits in a
transmission line resonator (the so called "transmons").
We demonstrate that the sensitivity of the optimized gate to $1/f$ flux and 
critical current noise is suppressed to leading order. 
The procedure only requires a preliminary estimate of the $1/f$ noise
amplitudes. No additional control or bias line beyond those 
used for the manipulation of individual qubits are needed.
The proposed optimization is effective also in the presence 
of relaxation processes and of spontaneous emission through the resonator (Purcell effect). 
\end{abstract}                                                                 


\maketitle

\section{Introduction}

Superconducting circuits are a promising technology for the
realization of quantum information on a solid state platform.
Several types of qubits~\cite{Clarke2008} have been developed realizing 
high fidelity single qubit operations~\cite{single-super,vion}. 
Rapid progress has also been made towards the realization of robust and 
scalable universal two-qubit gates~\cite{two-qb-exp,Bell-generation,Bell-violation}.  
The circuit quantum electrodynamics (cQED)~\cite{Blais2004}
architecture demonstrated to be particularly promising for scalable
quantum information.
In this scheme highly entangled two~\cite{two-cQED,Dewes1} and three 
qubits~\cite{DiCarlo10} have been generated and simple
quantum algorithms have been demonstrated~\cite{Dewes2,DiCarlo09}.

The coherence times of the present generation of devices ($\sim \mu$s) are about
three orders of magnitudes larger than the first implementations. A relevant step 
further toward this enhancement has been the elimination of linear sensitivity to 
low-frequency ($1/f$) noise by operating qubits at "optimal" working points. After 
the first "sweet spot" operation demonstrated in Ref.~\cite{vion}, a further boost 
of qubit performances has been achieved in a cQED design named "transmon"~\cite{Koch2007}, 
which is almost insensitive to the detrimental effect of $1/f$ charge 
noise~\cite{Schreier2008} at the price of  reduced anharmonicity.  However, cQED 
architectures share with other implementations the presence of $1/f$ flux noise whose 
amplitude has a characteristic order of magnitude~\cite{Wellstood}, and of $1/f$ critical 
current noise~\cite{Koch2007}. Together with relaxation processes due to quantum noise,   
dephasing due to 1/f flux and critical current noise still limits 
the time scales over which   phase coherence and {\em entanglement} are preserved.
In fact, further improvement of the coherence times at least of one order of magnitude 
would be required to reach the level for practical quantum error correction~\cite{Preskill}. 
Recently in a new circuit-QED architecture employing a three-dimensional resonator
the error correction threshold has been approached~\cite{Paik2011}. 
"Optimization" is thus a key-word of the present generation of superconducting nano-circuits. 
Clever circuit design and optimal tuning of multi-qubit architectures, supplemented by the 
use of improved materials, are two complementary strategies currently exploited to address 
this problem.

A major question currently unsolved is establishing the best strategy to maintain long-enough 
a sufficient degree of entanglement.
In the present article we address this issue considering a universal two-qubit gate realized 
by a fixed capacitive coupling of two transmons in a cQED architecture. 
The implementation of this scheme has been recently reported in Ref.~\cite{Dewes1}
where a  $\sqrt{\rm{i-SWAP}}$ operation with individual single-shot non-destructive
readout~\cite{Mallet2009} and gate fidelity of $90 \%$, partly
limited by qubit decoherence, has been demonstrated.
A similar system has been studied theoretically in ~\cite{Gywat,Rebic}.  
Here we identify "optimal"~\cite{Paladino2010} operating conditions of a transmons
 $\sqrt{\rm{i-SWAP}}$ gate
taking into account the multi-level nature of the nano-circuit. 
We find that an "optimal coupling" exists where the leading 
order effects of $1/f$ flux and critical current noise are eliminated. 
The amount of preserved entanglement is quantified by the  concurrence 
between the two transmons, $C(t)$, which we evaluate in analytic form.
The efficiency of the "optimal coupling" is demonstrated by the fact that, 
for typical $1/f$ noise spectra measured in superconducting nanocircuits,  
the concurrence is predicted to decay on a time scale~\cite{PaladinoNJP11}  
$T_2^{*\rm SWAP} \gtrsim 300 \, \mu$s (in the absence of other decay mechanisms). 
In addition, $C(t)$ may attain values~\cite{Verstaete} guaranteeing violation of a 
Bell inequality until $\sim 80 \, \mu$s and the gate fidelity is $99 \%$ up 
to $\sim 20 \mu$s. 
Finally, we demonstrate that the  optimization is effective also in the presence 
of relaxation processes due to flux quantum noise. 
Similarly to other cQED systems~\cite{Houck2008}, the gate efficiency can be limited
by spontaneous emission through the resonator. This limitation is likely to be
overcome by suitable Purcell filters or protected designs~\cite{Purcellprotection}.
The optimization proposed in the present article  can  further improve the considerable 
performance of cQED two-qubit gates based on cavity-mediated 
interaction~\cite{DiCarlo09,DiCarlo10} or on tunable effective interaction with microwave 
control~\cite{Chow2011}.
Remarkably, here effective elimination of omnipresent $1/f$ noise sources
is achieved even if one qubit does not operate at optimal bias and without 
additional controls or bias lines beyond those 
used for the manipulation of individual qubits, an important feature for scalability.

\section{Universal two-transmon gate}
\begin{figure}
\begin{center}
\includegraphics[width=0.8\textwidth]{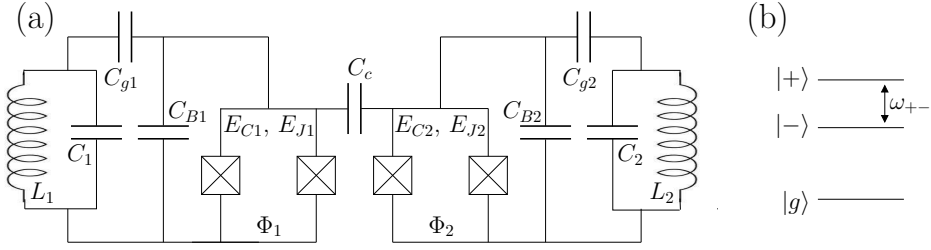}
\end{center}
\caption{(a) Circuit diagram of the two transmon gate with capacitive 
coupling energy $E_{CC}=(2e)^2 C_T/C_{\Sigma 1}C_{\Sigma 2}$, where
$1/C_T = 1/C_c+1/C_{\Sigma 1}+1/C_{\Sigma 2}$ and 
$C_{\Sigma \alpha}= C_{J \alpha} + C_{g \alpha} + C_{B
\alpha}$. (b) Schematic level structure.}
\label{fig1}
\end{figure}

We consider two transmons with a fixed capacitive coupling, each qubit being embedded in its 
superconducting resonator used for control and bit-wise readout~\cite{Dewes1,Mallet2009}. 
The interaction is effectively switched on/off by dynamically changing the qubits detuning using 
single qubits control lines. For this reason one of the qubits does not operate at
its sweet spot. In \fref{fig1} (a) we report the circuit diagram of the considered system. Each
transmon, denoted by the subfix $\alpha=1,2$, consists of a Cooper-Pair-Box (CPB) characterised 
by the charging energy  $E_{C \alpha}$ and Josephson energy $E_{J \alpha}= E_{J\alpha}^0 \cos(\phi_\alpha)$,
tunable via the magnetic flux threading the superconducting loop, $\phi_\alpha= \pi \Phi_\alpha/\Phi_0$, 
($\Phi_0$ is the flux quantum). In the circuit-QED scheme each CPB is embedded in a transmission
line resonator whose relevant mode is modeled as a $LC$ oscillator~\cite{Koch2007}. 
Thus the Hamiltonian of transmon $\alpha$ consists of the  CPB Hamiltonian plus
the dipole-like interaction with the $LC$ oscillator~\cite{Koch2007}
\begin{equation}
\!\!\!\!\!\!\!\!\!\!{\mathcal H}_\alpha = E_{C \alpha} (\hat q_\alpha - q_{x, \alpha})^2 - E_{J \alpha}(\phi_\alpha) 
\cos \hat \varphi_\alpha 
+ \omega_{r \alpha} \,  a_\alpha^\dagger a_\alpha + 2 \beta_\alpha e V_\alpha \, \hat q_\alpha \,
(a_\alpha + a_\alpha^\dagger) \, ,
\label{transmon}
\end{equation}
where phase, $\hat \varphi_\alpha$ and charge, $\hat q_\alpha$,  are conjugate variables, 
$[\hat \varphi_\alpha,\hat q_\alpha]=i$.
The resonator energy is $\omega_{r \alpha}= 1/\sqrt{L_\alpha C_\alpha}$  and $a_\alpha^\dagger$
($ a_\alpha$) creates (annihilates) one photon in the transmission line ($\hbar=1$).
$V_\alpha = \sqrt{\omega_{r \alpha}/2 C_\alpha }$ is the root-mean-square voltage of 
the oscillator and $\beta_\alpha= C_{g \alpha}/C_{\Sigma \alpha}$ is
the ratio between the gate capacitance coupling the CPB to the local mode and the CPB total
capacitance.  

The transmon operates at $E_{C \alpha} \ll E_{J \alpha}^0$.
Under these conditions, values of the phases $\varphi_\alpha$ close to zero are most favored. 
This motivates the  neglect of the periodic boundary condition on the phases  and the
expansion of the cosine in Eq. (\ref{transmon}). 
Within this approximation, the offset charge $q_{x, \alpha}$ can
be eliminated via a gauge transformation~\cite{Koch2007}. 
Of course, the perturbative scheme cannot capture the non-vanishing charge dispersion of the 
transmon~\cite{Catelani2011,Koch2009}.
In particular,  the exponential decrease of the charge dispersion with $\sqrt{E_{J \alpha}^0/E_{C \alpha}}$ 
for  $E_{J \alpha}^0/E_{C \alpha} \gg 1$ only results from the exact diagonalization of the CPB 
Hamiltonian in the phase basis~\cite{Koch2007}. It leads to 
exponential suppression of sensitivity to low-frequency ($1/f$) fluctuactions of the offset 
charge~\cite{Schreier2008}.
Here we rely on this well established result and eliminate $q_{x, \alpha}$ from the outset.
Expanding the cosine in Eq. (\ref{transmon}) up to fourth order, the CPB Hamiltonian
can be cast in the form of a weakly anharmonic oscillator (Duffin oscillator)
\begin{equation}
{\mathcal H}_{\alpha}^{\rm D} = \Omega_\alpha \,  b_\alpha^\dagger b_\alpha 
- (E_{C \alpha}/48)(b_\alpha + b_\alpha^\dagger)^4 \, ,
\label{Duffin}
\end{equation}
where the bosonic operators $b_\alpha$, $b_\alpha^\dagger$ are related to
the charge operator via 
$\hat q_\alpha = -i (E_{J \alpha}/2E_{C \alpha})^{1/4} (b_\alpha - b_\alpha^\dagger)/\sqrt 2$
and we put $\Omega_\alpha \equiv \Omega_\alpha(\phi_\alpha) = 
\sqrt{2 E_{C \alpha} E_{J \alpha}(\phi_\alpha)}$.
The two lowest eigenenergies of ${\mathcal H}_{\alpha}^{\rm D}$ identify the transmon-$\alpha$ qubit
levels. Their splitting is  $\tilde \Omega_\alpha = \Omega_\alpha - E_{C \alpha}/4$ and it can
be tuned by changing the magnetic flux $\Phi_\alpha$. The flux "sweet-spot" is at 
$\phi_\alpha=0$~\cite{vion,Koch2007}. 

The capacitive coupling between the CPBs, $E_{CC}(\hat q_1 - q_{x, 1})(\hat q_2 - q_{x, 2})$,
adds to $\sum_\alpha {\mathcal H}_\alpha$ leading to the Hamiltonian
\begin{equation}
{\mathcal H} = {\mathcal H}_{1}^{\rm D} +  {\mathcal H}_{2}^{\rm D} + 
\frac{\bar E_{CC}}{2} \, (b_1 - b_1^\dagger) (b_2 - b_2^\dagger) \, ,
\label{trasmoSWAP}
\end{equation}
where  $\bar E_{CC} = E_{CC} (E_{J1}E_{J2}/4 E_{C1} E_{C2})^{1/4} $ is the effective coupling 
depending on the control parameters $\phi_\alpha$ via the Josephson energies. 
Note that fluctuations of the magnetic fluxes affect the effective coupling between the qubits.
Typical values~\cite{Dewes1} are $E_{C \alpha} \sim 1$~GHz and $E_{J \alpha}^0 \sim 20-30$~GHz,
leading to $\bar E_{CC} = 10^{-1} - 10^{-2} $GHz.

The coupled transmons eigenenergies and eigenstates are conveniently  obtained by treating 
in perturbation theory with respect to $\sum_\alpha \Omega_\alpha \,  b_\alpha^\dagger b_\alpha $ 
both the anharmonic terms and the capacitive interaction included in
\begin{equation}
V= \frac{\bar E_{CC}}{2}  (b_1 - b_1^\dagger) (b_2 - b_2^\dagger) - 
\sum_\alpha  \frac{E_{C \alpha}}{48}(b_\alpha + b_\alpha^\dagger)^4 \, .
\end{equation}
The level structure is schematically show in \fref{fig1} (b). The splitting in the subspace where
the SWAP operation takes place (in short "SWAP splitting") reads 
$\omega_{+-}= \sqrt{(\tilde \Omega_1 - \tilde \Omega_2)^2 + \bar E_{CC}^2}$ and
the corresponding eigenstates spanning the ''SWAP subspace'' are $|- \rangle = - \sin (\eta/2) |01 \rangle + \cos (\eta/2) |10 \rangle$
and $|+ \rangle = \cos (\eta/2)  |01 \rangle +  \sin (\eta/2) |10 \rangle$, where
$\tan \eta =\bar E_{CC}/(\tilde \Omega_1 - \tilde \Omega_2)$ and
$|a,b \rangle \equiv |a \rangle_1 |b\rangle_2$ 
are eigenstates of $\sum_\alpha \Omega_\alpha \,  b_\alpha^\dagger b_\alpha $,
($a$, $b \in \{0,1\}$).
The interaction is effectively switched on  by tuning the single-qubit energy spacing 
to  mutual resonance. The resonance condition is realized by tuning the flux bias until 
$\tilde \Omega_1= \tilde \Omega_2$, displacing one qubit from the sweet spot at $\phi_\alpha=0$. 
In the following we suppose that $\phi_1=0$ and $\phi_2 \neq 0$. Under resonance conditions the 
$\sqrt{{\rm i-SWAP}}$ operation  $| 01 \rangle \to |\psi_e \rangle = [| 01 \rangle - i | 10 \rangle]/\sqrt{2}$
is realized by free evolution for a time  $t_E = \pi/2 \omega_{+-}$ starting from a 
factorized initial state in the ''SWAP subspace''.

\section{Optimal operating conditions: reduction of $1/f$ noise effects} 
\label{optimal}

\begin{figure}
\begin{center}
\includegraphics[width=0.9\textwidth]{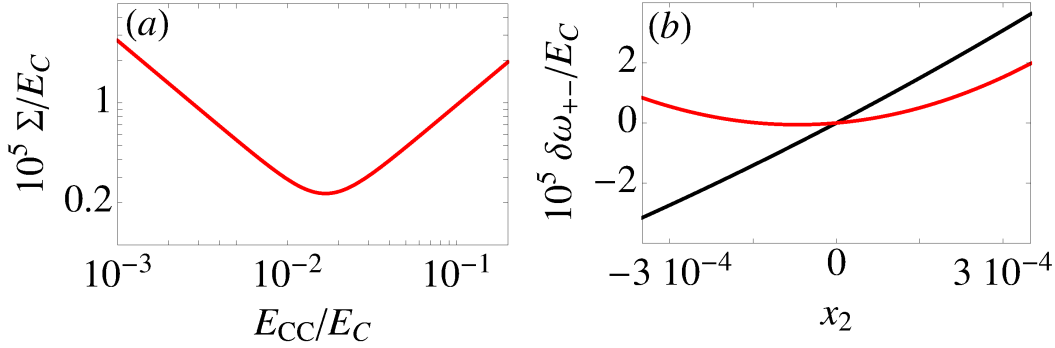}
\end{center}
\caption{Panel(a):  SWAP-splitting variance Eq.(\ref{variance})
as a function of  $E_{CC}/E_{C}$. The value of the minimum of
$\Sigma$ at the optimal point 
$E_{CC}^{\rm opt}=2 E_C  (\Sigma_{x_2}/\sqrt{2})^{1/2} = 1.68\cdot10^{-2} E_{C} $   
is one order of magnitude smaller than at $E_{CC}= 10^{-1}E_{C}$. 
Panel (b): Dispersion branch 
$\delta \omega_{+-}(x_2,x_1=0)/E_{C}$  for $|x_2| \leq 3 \Sigma_{x_2}$.
The black line is for a generic coupling $E_{CC}= 10^{-1}E_{C}$, the
red line is for the optimal coupling $E_{CC}^{\rm opt}$. 
Parameters are $E_{C \alpha}= 1$GHz, $E_{J \alpha}^0= 30 E_{C \alpha}$ with $\phi_2=0.64$ and
 $\Sigma_{x_2}= 10^{-4}$.}
\label{fig2}
\end{figure}

Since the two qubits do not operate at the same working point, the dominant
source of dephasing is different for the two transmons.
In particular, first order fluctuactions of the transmon splittings
are due to $1/f$ critical current noise for transmon 1
and to $1/f$ flux noise for transmon 2~\cite{Koch2007}.
These fluctuations can be treated in the adiabatic and longitudinal approximation~\cite{Falci}
by replacing $E_{J \alpha}$  with $E_{J \alpha}(1+ x_\alpha(t))$. Here $x_\alpha(t)$
represent stochastic fluctuations of the dimensionless critical current  
$x_1(t) = \delta I_{c1}(t)= \Delta I_{c1}(t)/ I_{c1}$,  
and of the flux $\Phi_2$, $x_2(t) = \tan(\phi_2) \, \delta \phi_2(t)$.
The leading order effect of adiabatic noise is defocusing, expressed by the
"static path" or static noise approximation (SPA)~\cite{Falci,Ithier} 
describing the average of signals oscillating at randomly distributed effective 
frequencies (see \ref{appendix-SPA} for the validity regimes of the SPA in the present problem). 
It is obtained by replacing $x_\alpha(t)$ with statistically
distributed values $x_\alpha$.  In the SPA the coherence 
between the states $|\pm \rangle$ is
\begin{equation}
\langle \rho_{+-}(t)  \rangle  \, = \, \rho_{+-}(0) \, e^{-i \omega_{+-}t } \, 
\langle e^{-i  \delta \omega_{+-} t}\rangle \, ,
\label{SPA-formal}
\end{equation}
where  $\rho(t)$ is the two-qubit density matrix  and $\langle ... \rangle$  indicates the average
over the fluctuations $x_\alpha$. 
Here we  assume that they are uncorrelated random variables 
with Gaussian distribution, zero mean and standard 
deviations $\Sigma_{x_\alpha}$ proportional to the amplitude of
the $1/f$ spectrum,   $S_{x_\alpha}^{1/f}(\omega) =\pi \Sigma_{x_\alpha}^2
[\ln(\gamma_{M \alpha}/\gamma_{m \alpha}) \, \omega]^{-1}$ ($\gamma_{m \alpha}$ and 
$\gamma_{M \alpha}$ are the low and the high frequency cut-offs of the $1/f$ 
region).
As demonstrated in Refs.~\cite{Paladino2010,PaladinoNJP11} 
the optimal operating condition is obtained imposing a minimum of the variance 
of the stochastic SWAP splitting,
$\Sigma^2 = \langle \omega_{+-}^2 \rangle - \langle \omega_{+-} \rangle^2$. 
This is simply understood considering the short-times expansion 
$|\langle  e^{- i \delta \omega_{+-} t}  \rangle | \approx \sqrt{1- (\Sigma t)^2}$,
implying defocusing suppression when $\Sigma$ is minimal. 
Expanding $\omega_{+-}$ around the fixed working point we get
\begin{equation}
\Sigma^2  \approx \sum_\alpha \left(
\frac{\partial \omega_{+-}}{\partial x_\alpha} \right)^2 \Sigma_{x_\alpha}^2 + \frac{1}{2}
\sum_{\alpha,\beta} \left(
\frac{\partial^2 \omega_{+-}}{\partial x_\alpha \partial x_\beta} \right)^2 
\Sigma_{x_\alpha}^2 \Sigma_{x_\beta}^2 \,,
\label{variance}
\end{equation}
where all derivatives are evaluated at $x_\alpha \equiv 0$.
At resonance  we find
$\partial \omega_{+-}/\partial x_\alpha= \bar E_{CC}/4$,
$\partial^2 \omega_{+-}/\partial x_\alpha^2 = - 3\bar E_{CC}/16 + \Omega^2/(4 \bar E_{CC}) $,
$\partial^2 \omega_{+-}/\partial x_1 \partial x_2= - \bar E_{CC}/16 - \Omega^2/(4 \bar E_{CC})$,
where we put $ \Omega_\alpha\equiv \Omega$, $E_{C\alpha}\equiv E_C$.
The variance Eq.~(\ref{variance}) is non-monotonic in the coupling energy (\fref{fig2} (a))
and its minimum depends on the noise  variances $\Sigma_{x_\alpha}^2$.
For typical values of the amplitudes of $1/f$ flux and critical current 
noise~\cite{nota2} the dominant effect is due to flux noise,
$\Sigma_{x_2} \gg \Sigma_{x_1}$ and the
optimal coupling is found at $ E_{CC}^{\rm opt} \approx 2 E_C  (\Sigma_{x_2}/\sqrt{2})^{1/2}$.
Note that, since $\bar E_{CC}$ depends on $x_\alpha$ (via $E_{J \alpha}$), 
the  differential dispersion $\partial \omega_{+-}/\partial x_\alpha$  at $x_\alpha=0$
is non vanishing unless the coupling is switched off. 
The condition of minimal variance effectively identifies an "optimal" dispersion 
leading to minimal defocusing, see \fref{fig2}(b). 

In addition we observe that, since $ E_{CC}^{\rm opt}$  depends on $E_C$ but not on the Josephson 
energy, the optimized SWAP frequency, $\omega_{+-}^{\rm opt} \approx \bar E_{CC}^{\rm opt}$, 
can be engineered by appropriately fixing (within the experimental tolerances) the 
ratios $E_{J \alpha}^0/E_{C \alpha}$.  This recipe can be conveniently applied even if an
independent estimate of the flux noise amplitude, $\Sigma_{x_2}$, for the specific setup is not 
available. In fact, the  variance  of the stochastic SWAP splitting, $\Sigma^2$, depends very
smoothly on $E_{CC}$ (\fref{fig2} (a)) allowing a practical estimate of $ E_{CC}^{\rm opt}$ 
based on the characteristic value of  $\Sigma_{x_2}$ observed in different flux and phase qubits. 
Alternatively, if different devices can be fabricated, one should select the sample with the ratio 
$E_{J \alpha}^0/E_{C \alpha}$ taking the right value for the given noise level 
of that particular device.

\begin{figure}
\begin{center}
\includegraphics[width=0.9\textwidth]{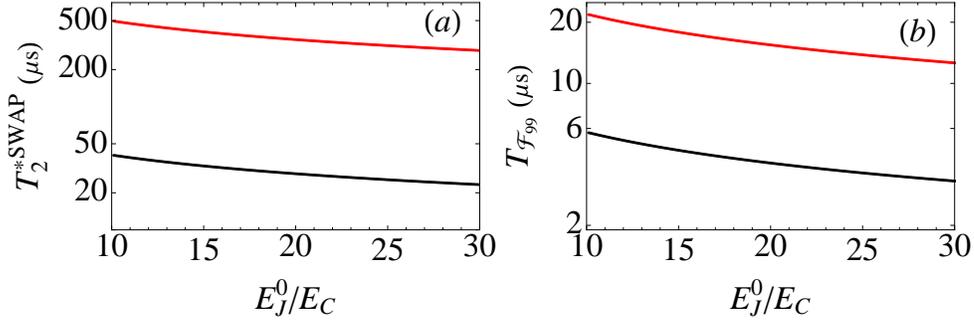}
\end{center}
\caption{$T_2^{*\rm SWAP}$ (panel a) and $T_{\mathcal F_{99}}$  (panel b) 
as a function of $E_J^0/E_C$ for $E_{C \alpha}= 1$GHz. 
The variances $\Sigma_{x_1}=5 \cdot 10^{-6}$, $\Sigma_{x_2}=10^{-4}$
correspond to typical values of $1/f$ critical current and flux noise~\cite{nota2}
rescaled to the present setup.
Black lines are obtained for  $E_{CC}= 10^{-1} E_{C}$, red lines correspond to
$E_{CC}^{\rm opt} =1.68\cdot10^{-2} E_{C}$. 
}
\label{fig3}
\end{figure}

The effectiveness against defocusing of  operating at the optimal coupling 
is revealed by the concurrence~\cite{concurrence}, which we evaluate in the SPA. 
We assume the system is prepared in the state $|01\rangle$ and freely evolves.
In the adiabatic approximation populations are constant thus
 $C(t) \approx 2 \,|{\rm Im} \langle \rho_{+-}(t) \rangle|$. Evaluating the integral
(\ref{SPA-formal}) we obtain 
\begin{equation}
C_{\rm SPA}(t) \approx 
 \Big  |{\rm Im}  
\frac{\exp{\Big\{- \frac{1}{2} \frac{\Sigma_{x2}^2 \left(
\frac{\partial \omega_{+-}}{\partial x_2} \right)^2 t^2}
{1+i \Sigma_{x2}^2 (\frac{\partial^2 \omega_{+-}}{\partial x_2^2}) t}\Big\}}}
{\sqrt{1+i \Sigma_{x2}^2 (\frac{\partial^2 \omega_{+-}}{\partial x_2^2}) t}} \Big | \, .
\label{concurrence}
\end{equation}
A measure of the entanglement preservation is the "SWAP decay time"~\cite{PaladinoNJP11} defined 
by the condition $|C_{\rm SPA}(T_2^{*\rm SWAP})| = e^{-1}$. 
At the optimal coupling $T_2^{*\rm SWAP}$ is one order of magnitude 
larger than for a generic coupling, assuming remarkable values up to
$T_2^{*\rm SWAP} \sim 400 \, \mu$s stable with increasing 
$E_J^0/E_C$,  \fref{fig3}(a).  In addition, a $99 \%$ fidelity to the Bell state $|\psi_{ent} \rangle=
[|01 \rangle + |10 \rangle]/\sqrt2$ is maintained up to 
$T_{\mathcal F_{99}} \approx 20 \,\mu$s, about $4$ times longer than for a generic
coupling, \fref{fig3}(b).  These results elucidate the capability of the proposed
operating condition to drastically reduce defocusing due to $1/f$ flux and critical
current noise. On the other hand, energy relaxation processes are expected to limit
the gate fidelity and the qubit relaxation times of the considered architecture~\cite{Dewes1}.
In the following Section we discuss the robustness of the optimal coupling condition to 
relaxation processes. 

\section{Optimal operating conditions: robustness to relaxation processes} 

We now discuss the robustness of the above optimization against relaxation processes due 
to flux noise and to spontaneous emission through the resonator.
Flux quantum noise is due to the external magnetic flux bias through a mutual inductance 
$M$~\cite{Koch2007} and it enters the Josephson energies in ${\mathcal H}_{\rm D}$ and in $\bar E_{CC}$.
It is included by adding to ${\mathcal H}$, Eq. (\ref{trasmoSWAP}),
the terms 
\begin{equation}
\Delta \mathcal H = - \frac{1}{2} \sum_\alpha \Omega_\alpha b^\dagger_\alpha b_\alpha \hat x_\alpha
- \frac{\bar E_{CC}}{4} \, (b_1 - b_1^\dagger) (b_2 - b_2^\dagger) \sum_\alpha \hat
x_\alpha \, .
\end{equation} 
For the transmon at the flux sweet spot it is $\hat x_1= \delta \hat \phi_1^2/2$, for 
transmon 2 instead $\hat x_2=  \tan \phi_2  \delta \hat \phi_2$,
where $\delta \hat \phi_\alpha$ are quantized phase fluctuations.
$\Delta \mathcal H$ conserves the parity of the total number of the two transmons excitations.
Thus it does not connect the states $|\pm \rangle$ to the ground state which, to the first order 
in $V$ takes the form
$| g \rangle \propto |00 \rangle + a_{11} |11 \rangle + \sum_{ij=1,2} a_{0j} |0, 2j  \rangle + a_{i0}
|2i, 0  \rangle $. 
Disregarding thermal excitation processes to higher energy states,
the only effect of flux quantum noise is inside the bi-dimensional subspace 
$\{ | \pm \rangle \}$. 
By solving a  Bloch-Redfield master equation~\cite{Cohen} 
relaxation and decoherence times  in the SWAP subspace are given  by the usual relation
$T_2^{\rm SWAP}= 2 T_1^{\rm SWAP}= \{ \frac{1}{16} 
\sum_\alpha (\sin \eta + (-1)^\alpha \cos \eta \bar E_{CC}/\Omega_\alpha )^2 S_{x_\alpha} (\omega_{+-})  \}^{-1}$
(a pure dephasing term $\propto S_{x_\alpha} (0)$ is disregarded with respect
to defocusing due to $1/f$ noise). 

\begin{figure}
\begin{center}
\includegraphics[width=0.6\textwidth]{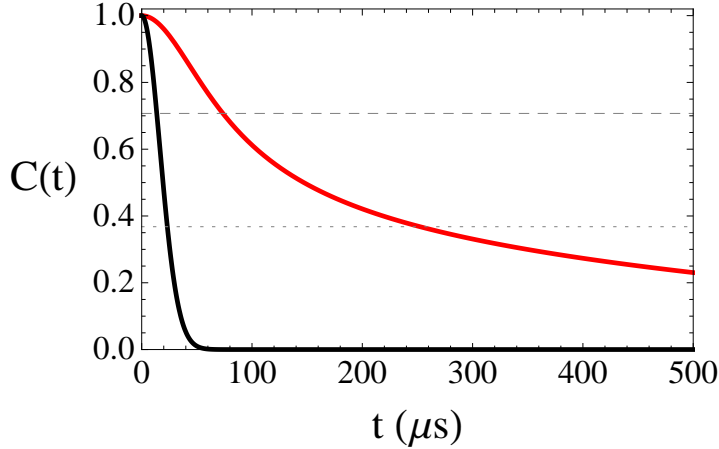}
\end{center}
\caption{Envelope of the concurrence in the presence of  $1/f$ flux 
and critical current noise (as in \fref{fig3}) and flux quantum noise 
on qubit 2 with spectrum $S_{x_2}(\omega)\simeq 10^{-9} \omega$.
Black line is for $E_{CC}= 10^{-1} E_{C1}$; the red line is for   
$E_{CC}^{\rm opt} =1.68\cdot10^{-2} E_{C1}$. The dashed gray line marks the value 
$C= 1/\sqrt{2}$, the dotted line 
$C(T_2^{\rm SWAP})= e^{-1}$.}
\label{fig:4}
\end{figure}

The main contribution comes from linear phase fluctuactions of the
transmon displaced from the sweet spot, $\hat x_2$. 
At low temperatures $k_B T \ll \Omega_\alpha$,  flux quantum noise  is
$S_{x_2} (\omega) \approx (\Omega_2 \tan \phi_2)^2 (\pi M/\Phi_0)^2 2 \omega/R$.
For typical parameters we estimate $T_2^{\rm SWAP} \approx 30$s at optimal coupling 
($M = 140 \Phi_0/A$, $R \sim 50 \Omega$)~\cite{Koch2007}. 
Thus the efficiency of the optimized gate on the SWAP time scale is not limited 
by relaxation processes due to flux noise.  
This is illustrated in \fref{fig:4} where we plot the {\em envelope} of the concurrence,
$C(t)$, which in the presence of $1/f$ noise and flux quantum noise reads
\begin{equation}
\!\!\!\!\!\!\!\!\!\!\!\!\!\!\!\!C(t) \approx \{ [\sin \eta (\rho_{++}(t) -\rho_{--}(t) ) + 
2 \cos \eta \, {\rm Re}\{\rho_{+-}(t)\}]^2  +  4  \,{\rm Im}\{\rho_{+-}(t)\}^2 \}^{1/2}.
\end{equation} 
Here the  population difference in the SWAP subspace is 
$\rho_{++}(t) -\rho_{--}(t) = (\cos \eta - \delta_{eq})e^{-t/T_1^{\rm SWAP}}
+  \delta_{eq}$, with $\delta_{eq}$ the thermal equilibrium value, and
 $\rho_{+-}(t) \equiv \langle \rho_{+-}(t) \rangle e^{-t/T_2^{\rm SWAP}}$.
We observe that for optimal coupling Bell inequality violation, guaranteed 
until~\cite{verstraete2002PRL} $C(t) \geq 2^{-1/2}$, 
occurs for times $\sim 75 \, \mu$s, much longer than for generic coupling.

In the cQED architecture each transmon is dispersively coupled to a resonator 
used for control and readout. An important mechanism for $T_1$ processes is spontaneous 
emission through the resonator (Purcell effect)~\cite{Koch2007,Houck2008}. 
If each transmon operates at positive resonator-transmon detunings, 
$\Delta_\alpha = \omega_{r \alpha} - \Omega_\alpha \sim 500 {\rm MHz} \gg g_\alpha \sim 50$~MHz,
where $g_\alpha$ is the transmon-resonator coupling strength, 
the spontaneous emission rate of  the coupled transmons 
is due to the "single-mode" Purcell effect~\cite{Houck2008,Mallet2009}. 
To evaluate it we rewrite ${\mathcal H}$, 
Eq. (\ref{trasmoSWAP}), in the basis of its perturbative eigenstates and
perform the rotating wave approximation eliminating terms describing the simultaneous excitation
(de-excitation) of one resonator and the coupled-transmons system. The restriction to the
subspace $\{ |g \rangle, | \sigma=\pm \rangle\}$ reads
\begin{equation}
 {\mathcal H}
+ \sum_\alpha \omega_{r \alpha} \,  a_\alpha^\dagger a_\alpha 
+ \sum_{\alpha, \sigma=\pm} (g_{\alpha, \sigma} |g \rangle \langle \sigma|  a_\alpha^\dagger 
\, + \, {\rm h. c.}) 
\label{H-RWA}
\end{equation}
where $g_{\alpha, \sigma} =i \sqrt 2 \beta_\alpha e V_\alpha (E_{J \alpha}/2E_{C \alpha})^{1/4}
\langle g | (b_\alpha - b_\alpha^\dagger) |\sigma \rangle$. The eigenstates of 
(\ref{H-RWA}) are obtained by treating the last term in first order perturbation 
theory. The ground state is unmodified and reads $|g, 0_1,0_2 \rangle$, where 
$|m_\alpha \rangle$ are Fock states of the $\alpha$-th resonator, $m_\alpha \in \bf N$. 
The corrections to the 
states $|\sigma,m_1,m_2 \rangle$ 
read $|\sigma,m_1,m_2 \rangle^{(1)} = \frac{g_{1 \sigma} \sqrt{m_1+1}}{E_\sigma - E_g - \omega_{r1}}
|g, m_1 +1,m_2 \rangle + \frac{g_{2 \sigma} \sqrt{m_2+1}}{E_\sigma - E_g - \omega_{r2}}
|g, m_1 ,m_2 +1 \rangle$, where $E_\sigma$ and $E_g$ are the unperturbed
eigenenergies of (\ref{H-RWA}).
The spontaneous decay rate is obtained applying  Fermi's
golden rule to  the interaction Hamiltonian of each resonator with
its harmonic bath. 
The transition rate from the coupled transmons plus resonators state 
$|\sigma,0_1,0_2 \rangle + |\sigma,0_1,0_2 \rangle^{(1)}$ to the ground state $|g, 0_1,0_2 \rangle$, 
is
\begin{equation}
w_\sigma =2 \sum_\alpha {\mathcal \kappa}_\alpha \Big | \frac{g_{\alpha, \sigma}}{E_\sigma - E_g - 
\omega_{r \alpha} } \Big |^2
\label{purcell}
\end{equation}
where ${\mathcal \kappa}_\alpha $ 
is the spontaneous emission rate of oscillator $\alpha$ 
and we considered single photon losses to each bath.
The coupled transmons SWAP levels experience a Purcell induced spontaneous 
emission rate reduced  with respect to the sum of the resonators spontaneous emission rates. 
For identical transmons in cavities with a lifetime $1/{\mathcal \kappa}_\alpha \approx 160$~ns 
we estimate $1/w_\sigma \approx 16 \mu$s (analogously to the transmon's relaxation time predicted in
Ref.~\cite{Koch2007}), signaling a limitation to the optimized gate efficiency.
We expect that the recently proposed Purcell filter or protection 
schemes~\cite{Purcellprotection} 
can be suitably extended to the considered two-qubit gate which is based on independent 
readout, possibly overcoming Purcell limitation. 

\section{Conclusions}
In conclusion, we demonstrated optimization of a cQED {\em entangling} gate 
against any relevant $1/f$ noise source while keeping the hardware simplicity of the 
fixed coupling and even if one qubit does not operate at optimal bias point. 
The estimated high performance of the gate signals the effective elimination of leading
order effects of $1/f$ noise.

Our analysis included all the relevant noise sources acting {\em during} the entanglement 
generating operation in the considered architecture. We have shown that the proposed scheme 
is robust with respect to relaxation processes due to quantum noise and it is likely to 
foresee a design protected also from Purcell effect. 
Additional errors during readout may of course influence the overall gate 
fidelity of any specific implementation~\cite{Dewes1}. The responsible error sources 
need to be independently eliminated.
However, the value of the optimal coupling is not affected by minimization of error sources
acting before/after the coupled-qubits evolution.
Similarly, for qubit-based quantum information~\cite{Nielsen}, optimization of single and two qubits 
quantum operations is a key requirement, even though the overall quantum processor will
suffer from error sources in between quantum operations or at preparation/readout.

Eliminating decoherence remains the biggest challenge for superconducting systems.
Further optimization may require on one side suppression of higher order effects of $1/f$ noise,
on the other limitation of relaxation due to quantum noise.  
Concerning intrinsic noise sources, like those responsible for $1/f$ noise, 
material engineering at the microscopic scale may be required in the near future. 
"Passive" optimization startegies may be conveniently combined with 
"active" control tools, like dynamical decoupling protocols inspired to nuclear 
magnetic resonance which have been already applied to superconducting systems~\cite{Bylander11,Falci2004}.
On a longer time scale, imperfections in the coherent control might represent the ultimate
limit to computer performance.

\ack
Discussions with G. Falci and information about ongoing experiments by D. Vion
are gratefully acknowledged. This work was partially supported by EU through 
Grant No. PITN-GA-2009-234970 and  by the Joint Italian-Japanese  
Laboratory on "Quantum Technologies: Information, Communication and  
Computation" of the Italian Ministry of Foreign Affairs.

\appendix
\section{Validity regime of the Static Path Approximation}
\label{appendix-SPA}

\begin{figure}[t!]
\begin{center}
\includegraphics[width=0.5\textwidth]{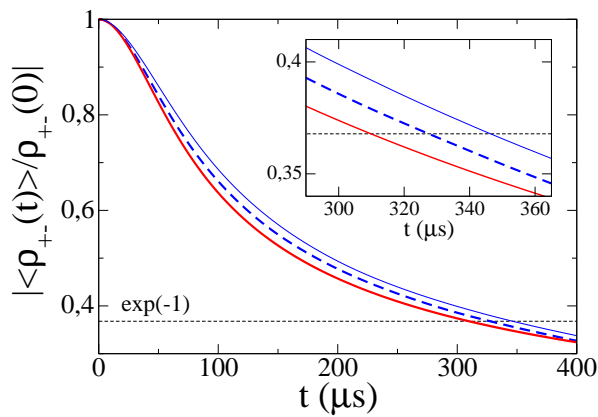}
\end{center}
\caption{Absolute value of the coherence $\langle \rho_{+-}(t)\rangle/\rho_{+-}(0)$ in the presence of  $1/f$ flux 
noise  on qubit 2 with  $\Sigma_{x_2}= 10^{-4}$.
The (thick) red line is the result of the SPA, the (thin) blue line is the 
numerical evaluation of the adiabatic approximation \eref{path-int}
for $\gamma_{m2} = 1 \, {\rm s}^{-1}$,  $\gamma_{M2}= 10^6 \, {\rm s}^{-1}$
as in \cite{Bylander11}, 
the dashed blue line is for $\gamma_{m2} = 1 \, {\rm s}^{-1}$, $\gamma_{M2}= 10^{5} \, {\rm s}^{-1}$:
the smaller is the high-frequency cut-off $\gamma_{M2}$, the closer is the SPA
to the adiabatic approximation.
Inset: zoom around the time range where $\rho_{+-}(t)/\rho_{+-}(0) \approx e^{-1}$.
Other parameters are
$E_{C \alpha}=1$~GHz, $E_{J \alpha}^0=30$~GHz, $E_{CC}=1.68 \times 10^7$~Hz 
($\bar E_{CC} \approx 65$~MHz). In the simulations we considered 
an ensemble of $\sim 10^3$ random telegraph noise processes with switching rates distributed 
as $\propto 1/\gamma$ in $[\gamma_{m2}, \gamma_{M2}]$. The average is performed using $10^6$ 
realizations of the stochastic process.}
\label{figureA1}
\end{figure}

The SPA is valid for times $t< 1/\gamma_{M \alpha}$. Thus it applies to the
considered $\sqrt{{\rm i-SWAP}}$ operation if 
$t_E= \pi/2 \omega_{+-} \approx 10^{-1} - 10^{-2} {\rm GHz}
 <  1/\gamma_{M \alpha}$. 
Since flux noise is the most relevant $1/f$ noise source in the considered setup,
here we disregard critical current fluctuations and consider the recent noise figures
reported in Ref.~\cite{Bylander11}. In that article
$1/f$ flux noise extends up to $\sim 1$~MHz, thus we can reasonably
expect that the condition $t_E < 1/\gamma_{M \alpha}$ is satisfied.

Moreover, we numerically  verified that the SPA is a valid approximation also for times 
$ t > 1/\gamma_M $ provided that $\gamma_M$ is smaller than the system oscillation 
frequency~\cite{Falci}.  In \fref{figureA1} we report the coherence between the states $|\pm \rangle$ in 
the SPA and the result of the numerical  evaluation of the adiabatic approximation
\begin{equation}
\rho_{+-}(t) = \rho_{+-}(0) \int \, \mathcal{D}[ \{ x_\alpha(s) \} ]  \, P[ \{ x_\alpha(s) \}] 
\, e^{-i \int_0^t ds \, \omega_{+-}(\{ x_\alpha(s) \})} \, ,
\label{path-int}
\end{equation} 
where $\omega_{+-}(\{x_\alpha(s) \} )\approx 
\sum_\alpha \frac{\partial \omega_{+-}}{\partial x_\alpha}  x_\alpha(s) + \frac{1}{2}
\sum_{\alpha, \beta}\frac{\partial^2 \omega_{+-}}{\partial x_\alpha \partial x_\beta}  
 x_\alpha(s) x_\beta(s) $ and the derivatives are reported in Section \ref{optimal} 
 below Eq. \eref{variance}. In the figure we considered flux noise on qubit 2, $x_2(s)$,
 distributed with $\Sigma_{x_2}= 10^{-4}$ and 
 $\gamma_{m2} = 1 \, {\rm s}^{-1}$,  $\gamma_{M2}= 10^6 \, {\rm s}^{-1}$~\cite{Bylander11}.
 It is clearly seen that the SPA is a reasonable approximation up to times
 $\sim 10^2/\gamma_{M2}$. 
 This legitimates the use of the SPA for the evaluation of the
 times  $T_2^{*\rm SWAP}$ reported in \fref{fig3}. The error with respect to an estimate 
 based on the adiabatic  approximation is of $\sim 10 \%$ (inset of \fref{figureA1}).

\end{document}